\documentclass[journal,transmag]{IEEEtran}

\usepackage{cite}
\usepackage{amsmath,amssymb,amsfonts}
\usepackage{algorithmic}
\usepackage{graphicx}
\usepackage{textcomp}
\usepackage{xcolor}
\usepackage{todonotes}
\usepackage{bm}
\usepackage{xspace}
\usepackage{subcaption}
\usepackage{booktabs}
\usepackage{tikzscale}
\usepackage{tikz}
\usepackage{pgfplots}
\usepackage{inputenc}
\usepackage{standalone}
\usepackage{multirow}
\usepackage{pdfpages}
\usepackage[section]{placeins}
\usepackage{float}
\usepackage{bm}
\usepackage[colorlinks, linkcolor={red!50!black},citecolor={blue!50!black},urlcolor={blue!80!black}]{hyperref}
\usepackage{setspace}
\usepackage{soul}
\usepackage{cancel}

\setstcolor{red}

\newcommand{\ybco}{YBa$_2$Cu$_3$O$_{7-x}$\xspace}
\newcommand{\grad}{\nabla}
\newcommand{\Href}{\textbf{H}$_{\textnormal{ref}}$\xspace}
\newcommand{\red}{\textcolor{red}}

\newlength\figH
\newlength\figW
\begin{document}
	%
	\title{Implementation of the H-$\phi$ formulation in COMSOL Multiphysics for Simulating the Magnetization of Bulk Superconductors and Comparison with the H-formulation}

	
	\author{\IEEEauthorblockN{Alexandre Arsenault\IEEEauthorrefmark{1},
			Fr\'ed\'eric Sirois\IEEEauthorrefmark{1}, and
			Francesco Grilli\IEEEauthorrefmark{2}}
		\IEEEauthorblockA{\IEEEauthorrefmark{1}Polytechnique Montr\'eal, Montr\'eal, Canada}
		\IEEEauthorblockA{\IEEEauthorrefmark{2}Karlsruhe Institute of Technology, Karlsruhe, Germany}
		\thanks{Corresponding author: Alexandre Arsenault (alexandre-1.arsenault@polymtl.ca)}
	}
	
	\IEEEtitleabstractindextext{%
		\begin{abstract}
			The H-formulation, used abundantly for the simulation of high temperature superconductors, has shown to be a very versatile and easily implementable way of modeling electromagnetic phenomena involving superconducting materials. However, the simulation of a full vector field in current-free domains unnecessarily adds degrees of freedom to the model, thereby increasing computation times. In this contribution, we implement the well known H-$\phi$ formulation in COMSOL Multiphysics in order to compare the numerical performance of the H and H-$\phi$ formulations in the context of computing the magnetization of bulk superconductors. We show that the H-$\phi$ formulation can reduce the number of degrees of freedom and computation times by nearly a factor of two for a given relative error. The accuracy of the magnetic fields obtained with both formulations are demonstrated to be similar. The computational benefits of the H-$\phi$ formulation are shown to far outweigh the added complexity of its implementation, especially in 3-D. Finally, we identify the ideal element orders for both H and H-$\phi$ formulations to be quartic in 2-D and cubic in 3-D, corresponding to the highest element orders implementable in COMSOL.
		\end{abstract}
		
		}

	\maketitle

	\IEEEdisplaynontitleabstractindextext

	%
	\IEEEpeerreviewmaketitle

	\section{Introduction}
	%
	%
	%
	%
	\IEEEPARstart{T}{he} modeling of the electromagnetic behavior of superconducting materials is a crucial aspect in the development of new technologies involving high temperature superconductors (HTSs). The simulation of magnetic fields produced by HTSs is especially important to predict and investigate the possible applications of HTSs without having to physically produce experiments, which require considerably more time and resources. Over the years, the Finite Element Method (FEM) has shown to be one of the most reliable computational methods to simulate electromagnetic fields of HTSs, with many different formulations adopted depending on the application of interest. 
	
	In the superconductivity community, the H-formulation, which uses a combination of Faraday's and Ampere's laws to solve for the magnetic field, has been widely used for the modeling of HTSs. Its applications vary remarkably, including the magnetization \cite{hong2006,ainslie2014,ainslie2015,philippe2015, zou2015, zou2016b, kapolka2018a, huang2020}, demagnetization \cite{celebi2015, kapolka2018, baghdadi2018} and magnetic levitation \cite{sass2015,grilli2018,queval2018, silva2019} of bulk superconductors, AC losses in superconducting windings \cite{brambilla2007,nguyen2011,zhang2012,ainslie2012,zermeno2013,xia2013, zhao2017,shen2020a}, etc. More details on the possibilities offered by the H-formulation can be found in two recently published review articles \cite{shen2020,shen2020a}.
	
	Brambilla et al. and J.P. Webb describe the many reasons for the appeal of this formulation \cite{brambilla2007,j.b.webb1993}. First, the use of edge elements allows the field to be discontinuous in the normal direction to the element edge, enabling the field to abruptly change direction near sharp corners. Additionally, the uniqueness of the solution for the magnetic field does not require any choice of gauge, as required by other formulations solving for the magnetic vector and/or scalar potentials. Moreover, the boundary conditions are easily implemented in this formulation due to the intuitive nature of the magnetic field as a dependent variable. Finally, the resulting dependent variables do not need extra calculations in order to obtain the magnetic field, as opposed to other formulations, such as the A-$\phi$ formulation, that must calculate spatial derivatives of \textbf{A} to obtain \textbf{H}. This last point is important because it reduces the numerical error when compared to other formulations, since computing the derivative introduces local inaccuracies. Note that the divergence-free condition of Maxwell's equations is not automatically enforced with curl elements, as previously assumed. The divergence-free condition is only met locally for elements of first order due to the discontinuities of the normal component of the field between elements \cite{zermeno2013a,wan2014}. However, in time-dependent simulations, the divergence-free condition is met at all times if the initial values are divergence-free \cite{zermeno2013}.
	
	Nevertheless, despite the renowned success of the H-formulation, this formulation still has its caveats. Namely, the solution of a vector field in the non-conducting regions, typically air, increases the size of the linear matrix to be solved. In reality, a full vector field is not required in current-free regions. Hence, although the H-formulation has been satisfactory for many applications, the computation times achieved using this formulation are longer than other formulations using nodal elements with the same mesh discretization\cite{lahtinen2015}. Furthermore, the H-formulation requires a dummy resistivity in the air regions, which is non-physical and degrades the matrix conditioning. Indeed, for the most accurate simulations carried out in 2-D in this work, the condition number of the stiffness matrix in the H-formulation reaches 2.58$\times10^{15}$, while it is merely 3.06$\times10^7$ in the H-$\phi$ formulation. The main reason for implementing the H-formulation everywhere in space is because it is simple to do, which explains why it is being used by over 45 research groups in the applied superconductivity community in the COMSOL Multiphysics finite element program \cite{shen2020,Comsol}. 
	
	On the other hand, mixed formulations such as H-$\phi$, T-$\phi$ and T-$\Omega$ are often used in electromagnetics dedicated FEM software \cite{carpenter1977,biro1995,zhou2008,Opera3d}. However, many of these software have their own shortcomings when applied to superconductivity: the non-linear resistivity cannot always be implemented depending on the software used and when it is possible, the convergence of the solution is not always favorable. Some researches have used the H-$\phi$ formulation to simulate superconductors in open-source software such as GetDP \cite{getdp,burger2019a,burger2019}, which provides greater customizability than COMSOL, but with a steeper learning curve and less support than commercial software.
	
	In this work, we implement the H-$\phi$ formulation in COMSOL Multiphysics 5.5 by coupling the magnetic field variables to the magnetic scalar potential through a custom-parameterized weak formulation. Surprisingly, such an implementation does not seem to have been reported yet despite its possibility since the earliest versions of COMSOL. The simulations conducted in this work have been tested on COMSOL version 4.3b and work as well as in the newest version of COMSOL. With a simple simulation of a 2-D and 3-D HTS bulk magnetized in a uniform applied field, we compare the computation times and accuracies between the H and H-$\phi$ formulations.

	\section{H-$\phi$ formulation description}
	
	\label{sec:H-phi}
	
\begin{figure}[tb!]
		\centering
		\includegraphics[width=\linewidth]{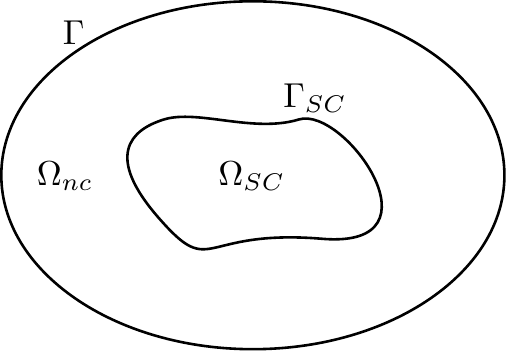}
		\caption{Arbitrary domain of simulation indicating the notation used in this work. Shown in the figure is the superconducting region ($\Omega_{SC}$), the non-conducting region ($\Omega_{nc}$), the boundary between these two domains ($\Gamma_{SC}$) and the external boundary of the air domain ($\Gamma$).}
		\label{fig:domain}
	\end{figure}

	The H-$\phi$ formulation combines the magnetic field, \textbf{H}, and magnetic scalar potential, $\phi$, in conducting and non-conducting regions, respectively. Based on Fig.~\ref{fig:domain}, the H-$\phi$ formulation can be stated as follows: one solves for the full vector $\mathbf{H}$ in the superconducting regions ($\Omega_{c}$, where currents can exist), and only for the magnetic scalar potential $\phi$ in the air regions ($\Omega_{nc}$, which are current-free). This allows for a reduction of the number of DOF in the problem. In this article, we refer to \textbf{H} and \textbf{h} as the magnetic fields in the H and $\phi$ physics, respectively. We can then derive $\mathbf{h}$ from $\phi$ as follows
	\begin{equation}
	\mathbf{h} = -\grad\phi \,.
	\label{eq:grad_phi}
	\end{equation}
	This definition forces $\mathbf{J}$ to be zero in the non-conducting regions since $\mathbf{J}=\nabla\times\mathbf{h}=\nabla\times\left(-\grad\mathbf{\phi}\right) = 0$ (the curl of a gradient is identically zero).
	
	To derive a useful equation from \eqref{eq:grad_phi}, one can use the divergence-free equation \hbox{$\nabla\cdot\mathbf{B}=0$}, and assume that \hbox{$\mathbf{B}=\mu_0\mathbf{h}$} in the whole simulated space containing the air and superconducting domains. We then obtain \hbox{$\nabla\cdot\left(\mu_0\mathbf{\mathbf{h}}\right)=\mu_0\nabla\cdot\left(-\grad\phi \right)=0$}, which can be rewritten as Laplace's equation, i.e.
	\begin{equation}
	\textrm{In $\Omega_{nc}$:}\quad \nabla\cdot\nabla\phi = 0 \,.
	\label{eq:phi}
	\end{equation}

	In the conducting regions, the governing equation is the conventional H-formulation expression, given by:
	\begin{equation}
	\textrm{In $\Omega_{c}$:}\quad \grad\times \left(\rho\grad\times\mathbf{H}\right) =-\mu_0\frac{\textrm{d} \mathbf{H}}{\textrm{d} t} \,,
	\label{eq:H}
	\end{equation}
	where the resistivity $\rho$ is nonlinear in the case of superconductors, therefore it cannot be taken out of the external curl operator. As mentioned earlier, when using the H-formulation in the whole simulation space, the resistivity of air must be set to a non-zero value in order to obtain better convergence. This problem does not exist when using the magnetic scalar potential in the air domains, since no resistivity must be specified.
	
	In order to generate a background magnetic field, we use a Dirichlet boundary condition in terms of $\phi$ on $\Gamma$. For example, if we want to apply a field in the +$\hat{\textnormal{y}}$ direction, the Dirichlet condition reads $\phi=-H_0(t)y$, with $H_0(t)$ being the time-dependent applied field. 
	
	Finally, we use curl edge elements in the conducting regions since we are solving for the vector \textbf{H}. We use Lagrange nodal elements in the non-conducting regions to solve for the scalar $\phi$. Knowing the properties of the elements used is important when coupling the two formulations, as described below. In the case of curl elements, the degrees of freedom are specified on the element edges, so that the tangential component of the dependent variable is constant along the element edges. On the other hand, Lagrange elements are nodal, meaning that the dependent variables are defined on element nodes and are free to vary along the element edges.
	
	\label{sec:couple}
	
	We first begin by coupling $\phi$  to the \textbf{H} variables. Since we have edge elements in the conducting domain, the DOF are given by the tangential component of the dependent variables on the element edges. Thus, we constrain the tangential component of \textbf{H} to be equal to the tangential component of \textbf{h} with the use of a constraint enforced in the strong form (called pointwise constraint in the Comsol language). This can be written as:
	\begin{equation}
	\hat{\mathbf{n}}\times \textnormal{\textbf{H}}=\hat{\mathbf{n}}\times \textnormal{\textbf{h}},
	\label{eq:tangent}
	\end{equation}
	where $\hat{\mathbf{n}}$ is the unit normal vector to $\Gamma_{SC}$.
	
	Coupling the \textbf{H} variables to $\phi$ is slightly more challenging. We first write~\eqref{eq:phi} in its weak form, which is easily shown to be:
	\begin{equation}
	\int_{\Omega_{nc}}\grad\phi\cdot\grad v\,dV-\int_{\Gamma_{SC}}\hat{\mathbf{n}}\cdot\grad\phi\,v\,dA=0\,,
	\label{eq:weakphi}
	\end{equation}
	where $v$ is a test function and $\mathbf{\hat{n}}$ is the vector normal to the $\Gamma_{SC}$ boundary. In this equation, the first term generates Gauss' law~($\grad\cdot\grad\phi=0$), while the second term allows fixing the normal flux of h~($=-\grad\phi$) on the $\Gamma_{SC}$ boundary.
	
	Since we have already constrained the tangential component of \textbf{H}, we must now couple the normal components in the weak formulation in order to completely define the physics at the interface between the H and $\phi$ domains. This is done by introducing the boundary condition $-\hat{\mathbf{n}}\cdot\grad\phi=\hat{\mathbf{n}}\cdot\mathbf{H}$ on $\Gamma_{SC}$, so that \eqref{eq:weakphi} becomes:
	\begin{equation}
	\int_{\Omega{nc}}\grad\phi\cdot\grad v\,dV+\int_{\Gamma_{SC}}\hat{\mathbf{n}}\cdot\mathbf{H}\,v\,dA=0\,.
	\label{eq:weakphi_couple}
	\end{equation}

	Instructions for modeling the equations of this section in COMSOL Multiphysics are detailed in the Appendix.
	
	\section{Magnetization simulations}
	
	We evaluate the computational efficiency of the H-$\phi$ formulation with a simple simulation of a bulk HTS magnetized by zero field cooling (ZFC) in a uniform background field of 5~T. The magnetic field is slowly ramped up and is brought back down using a smoothed triangular function of one second in duration. We use COMSOL Multiphysics 5.5 to solve the above equations with the finite element method. The personal computer used to perform the simulations in this paper has an Intel(R) Core(TM) i7-3770 processor with 32~Gb of random access memory. 
	
	\begin{figure}[tb!]
		\centering
		\includegraphics[width=\linewidth]{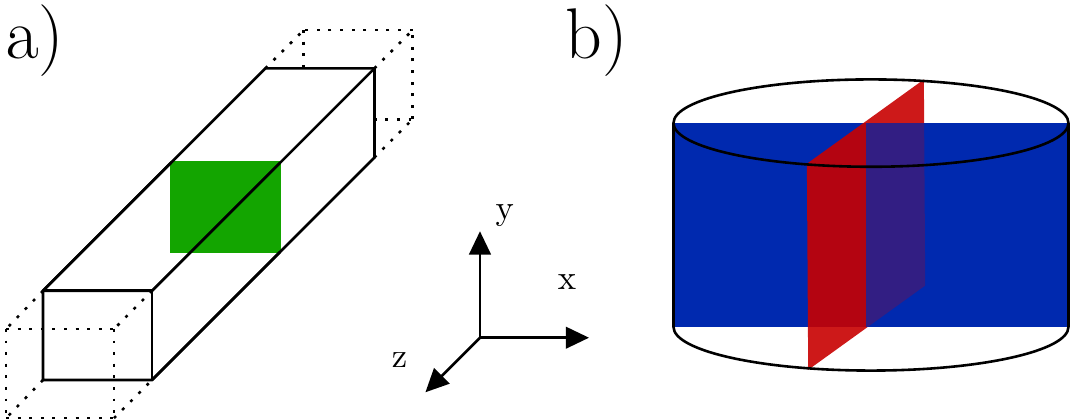}
		\caption{Superconducting bulks considered in the 2-D and 3-D simulations. a) An infinitely long superconducting bar simulated in 2-D by considering only the green cross section shown. b) A cylindrical superconducting bulk simulated in 3-D. The blue and red shaded planes represent the x-y and y-z planes used to visualize the resulting field, current density and local errors in subsequent figures.}
		\label{fig:geom}
	\end{figure}
	
\begin{figure}[tb!]
		\centering
		\includegraphics[width=\linewidth]{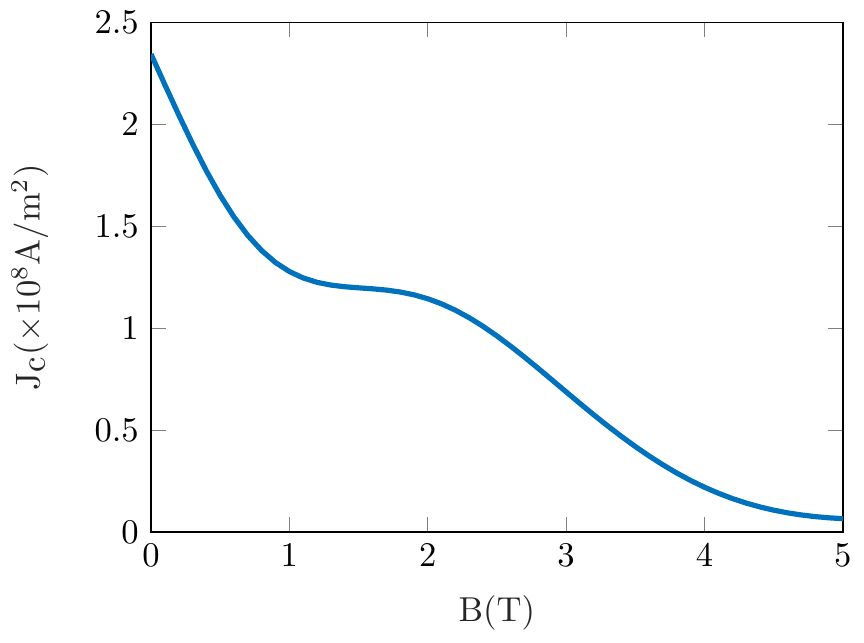}
		\caption{Smoothed function of the J$_{\textnormal{c}}(\|\mathbf{B}\|)$ data measured on a \ybco bulk by Can Superconductors \cite{superconductors}.}
		\label{fig:squid}
	\end{figure}

	The HTS geometry considered is an infinitely long bar in \mbox{2-D} and a cylindrical bulk in 3-D as shown in Fig.~\ref{fig:geom}a) and b), both of $2\times 1$~cm cross section. The non-linear resistivity of the HTS is modeled using the power law model \cite{rhyner1993}:
	\begin{equation}
	\rho=\frac{E_c}{J_c(\|\mathbf{B}\|)}\left(\frac{\|\mathbf{J}\|}{J_c(\|\mathbf{B}\|)}\right)^{n-1},
	\end{equation}
	with $\mathbf{J}$ being the current density, J$_{\textnormal{c}}(\|\mathbf{B}\|)$ being the field dependent critical current density, $n=25$, and $E_c=1~\mu$V/cm. We use Superconducting QUantum Interference Device (SQUID) measurements obtained by Can Superconductors \cite{superconductors} on a \ybco bulk as an input for modeling the field dependence of the critical current density at 77~K, as shown in Fig.~\ref{fig:squid}. The SQUID data is smoothed in order to obtain better convergence. We compare the results obtained from the H-$\phi$ formulation to those obtained from the H-formulation in both 2-D Cartesian and 3-D simulations.

	We use a MUltifrontal Massively Parallel sparse direct Solver (MUMPS) for the H-formulation and a PArallel Sparse DIrect Solver (PARDISO) for the H-$\phi$ formulation, since these solvers were found to give better computation times and accuracies for the respective formulations both in 2-D and 3-D. A relative tolerance of 1$\times10^{-6}$ is used in all the simulations.
	
	
	The geometry used to simulate the HTS bar in 2-D Cartesian form is shown in Fig.~\ref{fig:geom}a), with a cross-section of $2\times1$~cm. A circular air domain with 15~cm radius is also considered. We use a triangular mesh on the whole domain of simulation, with a mesh 5 times finer inside the superconducting domain than in the air domain.
	
\begin{figure}[tb!]
		\centering
		\includegraphics[width=0.99\linewidth]{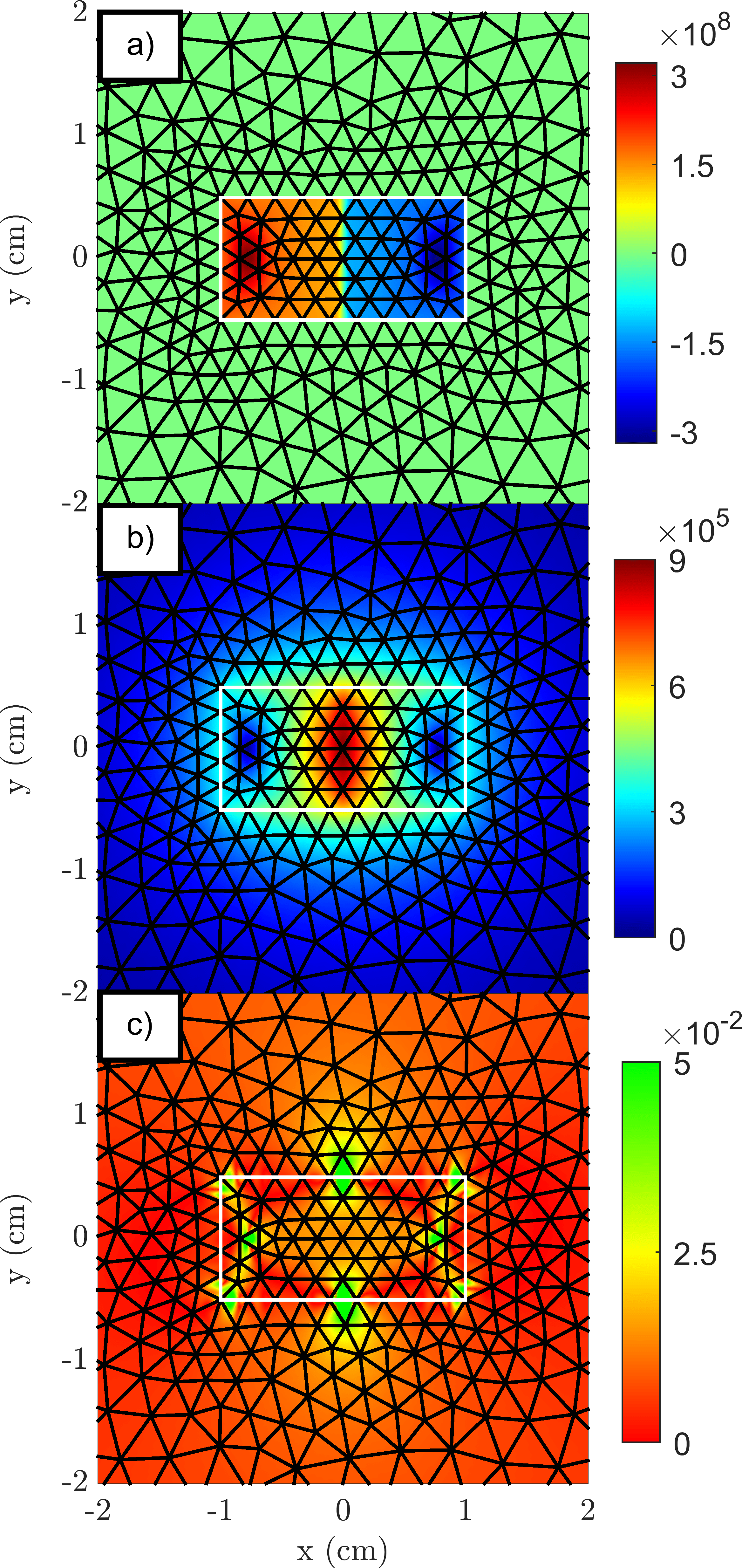}
		\caption{a) Current density in units of A/m$^2$ and b) norm of the magnetic field in units of A/m calculated using the H-formulation with 2594 quartic elements after the background field is ramped down to zero. This result is taken as the reference for comparison with other simulations, as explained in the text. c) Percent error of the magnetic field calculated at each point using the H-$\phi$ formulation, with the H-formulation as reference, both using 2594 quartic elements. The maximum value of the colorbar has been reduced from 0.41~\% to 0.05~\% in order to better visualize the location of the errors. The values corresponding to 0.41~\% are the two points inside the superconducting domain where the field is nearly zero. The error of the field near the edges of the bulk remains below 0.3~\%. The black lines represent the mesh.\vspace{-1cm}}
		\label{fig:Bref}
	\end{figure}

	In order to show that the magnetization is properly simulated, we present the current density distribution obtained after the background field is ramped down to zero in Fig.~\ref{fig:Bref}a). The maximum/minimum of the current density is not at the edges of the bulk due to the field dependent critical current density. This result is found using 2594 quartic elements with the 2-D H-formulation, but the H-$\phi$ formulation yields nearly identical results. Nonetheless, considering the dependent variables used in the simulations are the magnetic field components, we compare only the magnetic field between both formulations in this work.
	
	The norm of the magnetic field after the ZFC process is shown in Fig.~\ref{fig:Bref}b), where the result was calculated using 2594 quartic elements with the H-formulation. This solution constitutes our reference solution. We compare this result with the one obtained with 2594 quartic elements in the H-$\phi$ formulation. For this, we compute the relative error (in percent) between both results using
	\begin{equation}
	e=\left|\frac{\|\mathbf{H}\|-\|\mathbf{H}_{\textnormal{ref}}\|}{\|\mathbf{H}_{\textnormal{ref}}\|}\right|\times 100\%,
	\label{eq:error}
	\end{equation}
	where $\|\mathbf{H}_{\textnormal{ref}}\|$ is the norm of \textbf{H} in the reference solution obtained with the H-formulation. Throughout this work, we use the same order of curl and Lagrange elements for the H-$\phi$ formulation unless stated otherwise. We use COMSOL's \textit{Join} feature to calculate the error between different simulation results.
	
	The percent error between formulations is shown in Fig.~\ref{fig:Bref}c). Regions of maximal error occur where the \textbf{H}-field varies more drastically at the edges of the bulk, more specifically at the corners and along the center of the top and bottom edges. This induced error most likely comes from the connection of the normal \textbf{B} components across elements with curl and Lagrange shape functions. However, even at these points, the maximum error is below 0.3~\% when using 2594 quartic elements. This error increases when fewer and lower order elements are used. At distances far from the bulk, the fields are practically equivalent, with a percent error less that 0.01\%. 
	
	Finally, the worst error is found in the superconducting domain to be $\sim 0.41~\%$, most likely emerging from the fact that the norm of the field is nearly zero at these points, therefore blowing up the denominator in \eqref{eq:error}. Although we do not explicitly impose Gauss' law in the H physics in this work, we verified that forcing the divergence-free condition has no significant impact on the local error calculated even though the $\phi$ physics is always divergence-free. This demonstrates that Gauss' law is still obeyed in the time-dependent H-formulation even without explicitly imposing it.
	
	\subsection{2-D H-formulation}
	In order to properly compare our simulation results, we must determine an accurate representation of the magnetic field produced by the HTS. We begin by simulating the well documented H-formulation and use the standard FEM method of varying the number of DOF in the model to verify the convergence of the results. We calculate the average of the norm of the magnetic field inside the superconducting region after the ZFC process as our observable quantity to accomplish the convergence rate analysis. In the rest of the paper, we shall refer to this observable quantity as the ``convergence parameter". With increasingly finer mesh and element order, the solution is said to be converged when increasing the number of DOF does not significantly affect the convergence parameter value.
	
\begin{figure*}[tb!]
		\centering
		\includegraphics[width=0.49\linewidth]{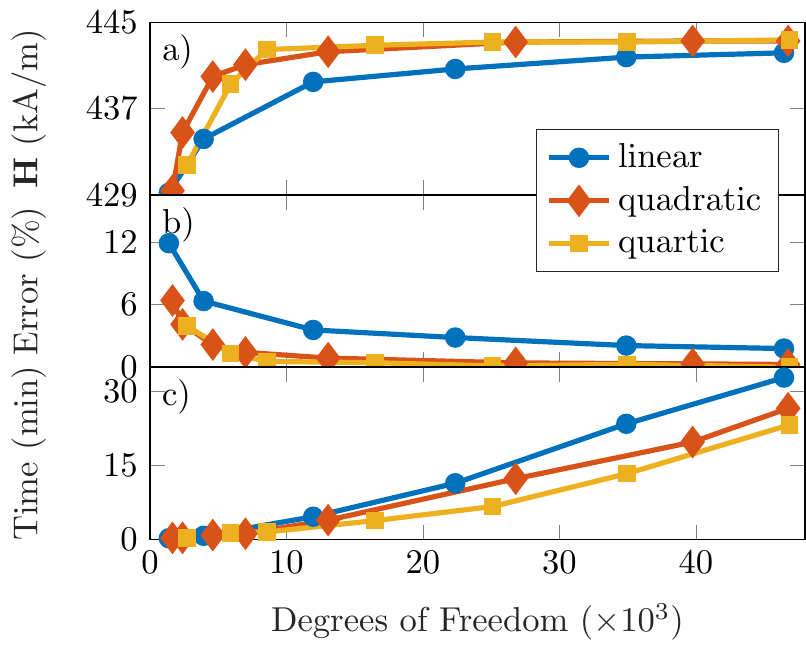}
		\includegraphics[width=0.49\linewidth]{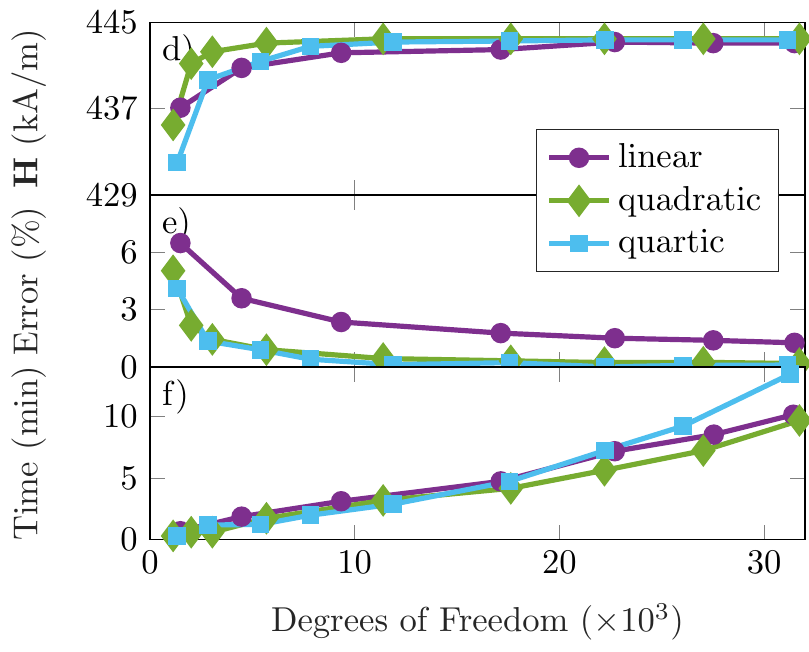}
		\caption{Comparison between 2-D simulations of the H and H-$\phi$ formulations with linear, quadratic and quartic elements.  a) and d) show the convergence rate of the average field over the superconducting domain as a function of DOF in the H and H-$\phi$ formulations, respectively. b) and e) show the percent error relative to 2594 quartic curl elements simulated in the H-formulation as a function of DOF in the H and H-$\phi$ formulations, respectively. Finally, computation times of the H (c) and H-$\phi$ (f) formulations as a function of DOF are shown. A MUMPS direct solver is used for the H-formulation, while a PARDISO solver is used for the H-$\phi$ formulation.}
		\label{fig:DOF_2D}
	\end{figure*}

	Fig.~\ref{fig:DOF_2D}a) summarizes the convergence rates of the average field with linear, quadratic and quartic elements using the H-formulation. According to the figure, quadratic elements with at least $\sim 13,000$ DOF and quartic elements with at least $\sim 9,000$ DOF are needed for a convergent result of $\sim 443$~kA/m. On the other hand, the solution barely converges even with the highest number of DOF for linear elements. 
	
	By referring to Fig.~\ref{fig:DOF_2D}a), we use 2594 quartic elements with 47,000 DOF as our ``exact" solution, shown in Fig.~\ref{fig:Bref}b). This solution will be used as a reference in order to compare with other simulation results in 2-D. The magnetic field calculated in this simulation will be referred to as \Href in the rest of this section.

	Fig.~\ref{fig:DOF_2D}b) illustrates the percent error calculated using different element orders and mesh discretizations for the H-formulation. The error is calculated by integrating \eqref{eq:error} over the superconducting domain and dividing by the area, while using \textbf{H}$_\textnormal{ref}$ as reference. The behavior is very similar to the convergence rates of Fig.~\ref{fig:DOF_2D}a). The percent error for linear elements remains above 1.7~\% even for the highest number of DOF, while the error for quadratic and quartic elements is negligible once the convergence of the norm of \textbf{H} is achieved. Note that quartic elements are at least 0.2\% more accurate than quadratic elements for a given number of DOF. 
	
	Finally, the computation times of each simulation in Fig.~\ref{fig:DOF_2D}a) are shown in Fig.~\ref{fig:DOF_2D}c). In addition to the lack of convergence and consistently higher percent error obtained with linear curl elements, the computation times are even higher than quadratic and quartic elements for a given number of DOF. For example, with $\sim46, 000$~DOF, linear elements are $\sim1.4$ times slower than quartic elements. This conclusion is surprising since most authors use first order edge elements with the H-formulation, which has been shown to help achieve good convergence in AC loss computations at power frequencies \cite{sirois2008}. However, the slow magnetization process simulated in this paper has not shown any convergence issues related to the order of the basis functions, which opens the door to further exploration. Nevertheless, in the scope of this paper, we can clearly state that quartic elements are favorable in 2-D, both for their computation times as well as their accuracy. Note that quartic elements are the highest element order available for curl elements in COMSOL in 2-D.
	
	\subsection{2-D H-$\phi$ formulation}
	
	We follow the procedure of the last section to determine the ideal element order of the H-$\phi$ formulation and compare the results with the H-formulation. 
	
	The convergence parameter as a function of DOF shows similar behavior as the one calculated in the H-formulation of Fig.~\ref{fig:DOF_2D}a), as seen in Fig.~\ref{fig:DOF_2D}d). However, for the H-$\phi$ formulation, the quadratic elements converge slightly faster than quartic elements. Additionally, linear elements converge above $\sim22,700$ DOF. Note that for a given mesh and shape function order, the number of DOF obtained in the H-$\phi$ formulation is less than half the amount obtained in the H-formulation. This difference depends on the size of air domains and will therefore vary depending on the model.
	
	We determine the accuracy of the H-$\phi$ formulation by calculating the percent error with \textbf{H}$_{\textnormal{ref}}$ as reference, presented in Fig.~\ref{fig:DOF_2D}e). We find that linear elements still do not accurately represent the exact solution, while quadratic and quartic elements have negligible error ($<0.45\%$) for more than $\sim12,000$ DOF. Comparing with the H-formulation errors of Fig.~\ref{fig:DOF_2D}b), the use of linear elements seems more forgiving in the H-$\phi$ formulation, since the error is systematically smaller in this case. However, the error obtained with quadratic and quartic elements are both very low, typically less than 1~\% for more than 5,000~DOF.
	
	Lastly, the computation times for the different element orders are compared in Fig.~\ref{fig:DOF_2D}f). We find that the computation times are similar for fewer DOF, while these times are $\sim30\%$ higher for quartic elements at the highest DOF calculated. We find that quadratic elements systematically yield less computation times than linear and quartic elements for the same number of DOF.
	
	Comparing Fig.~\ref{fig:DOF_2D}a), b) and c) with Fig.~\ref{fig:DOF_2D}d) e) and f), we find the main advantage of using the H-$\phi$ formulation. In the case of the H-$\phi$ simulations, the error becomes negligible for nearly half the amount of DOF than that of the H-formulation for quadratic and quartic elements. Furthermore, the computation times are nearly three times faster for the H-$\phi$ formulation for the most accurate solution obtained, representing a percent error of only 0.015~\%. The computation time difference is not as drastic for lower amounts of DOF, but the H-$\phi$ formulation is still more than twice as fast when a percent error less than $\sim15\%$ is required with quartic elements.
	
	For a final comparison, we consider the computational time required for each formulation and element order to obtain a relative error of 0.5\%, as shown in table~\ref{tbl:2D}. We find that quartic elements need the lowest amount of time to reach an error of 0.5\% in both formulations. While the improvement in computation time for the H-$\phi$ formulation is not substantial (35 seconds), the improvement will certainly increase for more complex models requiring more degrees of freedom.
	
	\begin{table}[h]
		\centering
		\caption{Computation times of the H and H-$\phi$ formulations in 2-D in order to achieve a relative error of 0.5\%}
		\begin{tabular}{ l  l  l }
			\toprule
			Order  & H time (min) & H-$\phi$ time (min)\\
			\midrule
			1  & $>$45 & $>$10 \\
			2  & 10.62 & 3.01 \\
			4  & 2.42 & 1.84 \\
			\bottomrule
		\end{tabular}
		\label{tbl:2D}
	\end{table}
	
	We end the 2-D analysis by studying the viability of using different shape function orders between the H and $\phi$ physics. In this case, we restrict the number of elements to 50 in the superconducting domain and vary the number of elements in the air domains in order to get 21,655$\pm$23 DOF for every combination of element orders considered.
		
	The results of the analysis are shown in table~\ref{tbl:mix2D}, where the computation times and percent errors are used for the comparison. In order to be more concise, we will refer to the combination of elements as (element order in H):(element order in $\phi$) for the rest of the article, so that using quadratic elements in H and linear elements in $\phi$ will be denoted 2:1. We find that mixed elements are always less accurate than their homogeneous counterparts despite requiring more computation time. 1:1 elements are $\sim$2.5~\% more accurate than 1:2 elements, even though 1:2 elements require 3~seconds more time. 2:2 elements are $\sim$0.5~\% more accurate than 2:4 elements and 47~seconds faster. In the end, 4:4 elements yield the most accurate results with only 0.098~\% error and a computation time 102~seconds faster than 4:6 elements. In order to determine if 4:4 elements are superior to 4:2 elements, we computed an additional 4:2 simulation with 30,434~DOF, since the computation time was faster than 4:4 elements when using 21,678~DOF. Even with additional elements and longer computation time, the error is still 0.325~\% greater with 4:2 elements than with 4:4 elements. We therefore conclude that 4:4 elements are still the best option within COMSOL limitations in 2-D. Note that similar conclusions in terms of the comparison of the performance between the two formulations, as achieved in 2-D Cartesian, could also be obtained in 2-D axisymmetric, since the two problems are formally coincident and all the properties discussed for H and $\phi$ apply.
	
	\begin{table}[h]
		\centering
		\caption{Mixed element order comparison for the 2-D H-$\phi$ formulation}
		\begin{tabular}{ l  l  l  l l}
			\toprule
			H Order & $\phi$ Order & DOF & Time~(min) & Error~(\%) \\
			\midrule
			
			\multirow{3}{*}{1} & 1 & 21,679 & 1.500 & 7.125\\
			{}	& 2 & 21,641 & 1.550 & 9.613\\
			{}	& 4 & 21,647 & 2.167 & 10.446\\\\
			
			\multirow{3}{*}{2} & 1 & 21,648 & 2.067 & 6.533\\
			{}	& 2 & 21,664 & 1.917 & 0.643\\
			{}	& 4 & 21,642 & 2.700 & 1.111\\\\
			
			\multirow{4}{*}{4} & 2 & 21,678 & 4.800 & 0.395\\
			{}	& 4 & 21,640 & 6.067 & 0.098\\
			{}	& 6 & 21,658 & 7.767 & 0.119\\
			{}	& 2 & 30,434 & 6.583 & 0.423\\

			\bottomrule
		\end{tabular}
		\label{tbl:mix2D}
	\end{table}
	
	\subsection{3-D H-formulation}
	
	The three-dimensional analysis is carried out by following the same procedure as in the 2-D case. This time, we simulate a cylindrical bulk of 1~cm radius and 1~cm height, with an air domain in the form of a sphere of 15~cm radius. The geometry of the superconducting domain is shown in Fig.~\ref{fig:geom}b). As in the 2-D case, the physics coupling of the H-$\phi$ formulation is done by using equations \eqref{eq:tangent} and \eqref{eq:weakphi_couple}. We apply the magnetic field along the z-axis, corresponding to the axial direction of the cylinder.

\begin{figure*}[tb!]
		\centering
		\includegraphics[width=1\linewidth]{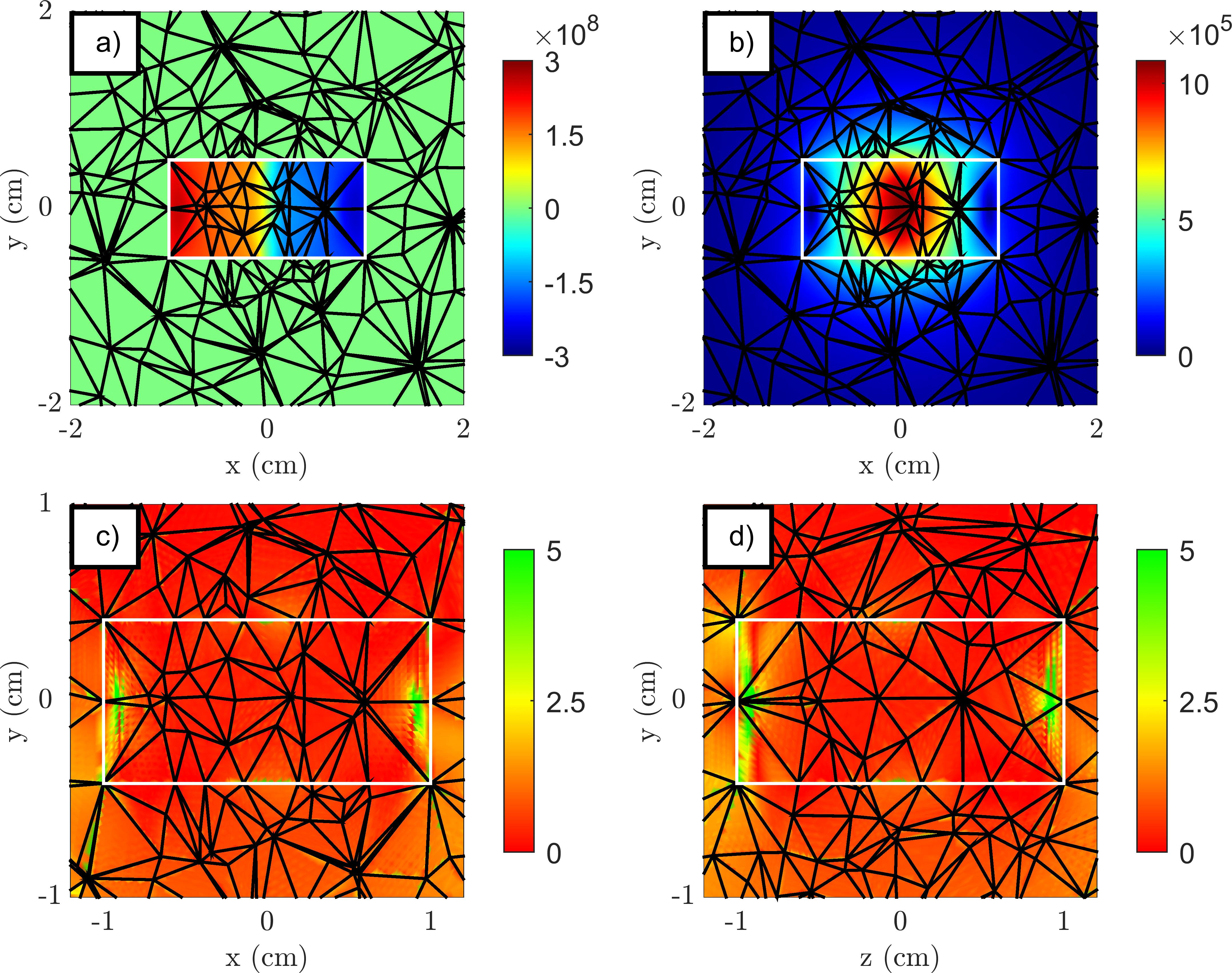}
		\caption{a) Current density in units of A/m$^2$ and b) norm of the magnetic field in units of A/m in the x-y plane calculated using the H-formulation with 8325 cubic elements after the background field is ramped down to zero. Also shown is the percent error of the magnetic field calculated at each point in the c) x-y and d) y-z plane using the H-$\phi$ formulation. The H-formulation is used as reference and both formulations use 8325 cubic elements. The maximum value of the colorbar has been reduced from 137~\% to 5~\% for clarity. The values corresponding to 137~\% are the two points inside the superconducting domain where the field is nearly zero. The error of the field near the edges of the bulk still remains below 20~\%. The white lines represent the superconductor/air boundary and the black lines represent the cross-section of the 3-D mesh on the 2-D plane. The axes have been adjusted in c) and d) to better visualize the local errors.}
		\label{fig:Bref3D}
	\end{figure*}

	The current density and reference field, \textbf{H}$_{\textnormal{ref}}$, used for the 3-D simulations are displayed in Fig.~\ref{fig:Bref3D}a) and b), respectively, where 8325 cubic elements are used in the H-formulation. The 3-D mesh is projected onto the x-y plane in order to illustrate the density of elements used. As expected from the H-formulation, we recover the typical behaviour of a superconducting bulk magnetized in a uniform field.
		
	The local error between the H and H-$\phi$ formulations on the x-y and y-z planes is shown in Fig.~\ref{fig:Bref3D}c) and d), respectively, where 8325 cubic elements are also used in the H-$\phi$ formulation. We reduced the maximum colorbar value from 137~\% to 5~\% in order to better perceive the local errors in regions of nonzero field. Similar to the 2-D case, the error is maximized in areas where the field approaches zero, corresponding to a divergence in \eqref{eq:error}. We again find that the error is more pronounced where the field increasingly varies, such as in the corners and the center of the top and bottom edges of the superconducting domain, reaching errors as high as 20~\% near the corners. However, the error is more concentrated near the element's edges in 3-D than in 2-D. In constrast to the 2-D case, there are still significant sources of error outside the superconducting domain concentrated near element edges. This is most definitely an artifact ensuing from slicing the 3-D geometry to calculate the error on the 2-D planes. Indeed, when the whole 3-D domain is considered, the errors of $\sim$5~\% located near element edges in the air domains vanish. We show only the 2-D cross section for the sake of clarity. Note that although the problem is inherently 2-D axisymmetric, the irregularities of the 3-D mesh produce slightly different local errors between the x-y and y-z planes.
	
	
	In order to obtain an accurate solution, we investigate the convergence of the solutions by varying the mesh discretization and element order with the H-formulation. The convergence parameter is obtained by calculating the average field over the superconducting domain. In the 3-D case, the highest available order of elements in COMSOL is cubic, so we compare the convergence rates of linear, quadratic and cubic elements, as illustrated in Fig.~\ref{fig:DOF_3D}a).
	
	Cubic elements converge slightly faster than quadratic elements, yet both element orders converge to the same value of $\sim$403~kA/m. Conversely, linear elements converge more slowly and to a value slightly higher than quadratic and cubic elements at $\sim$406~kA/m.
	
	We determine the percent error of each simulation with respect to the number of DOF using the solution with $157,400$ DOF and 8325 cubic elements as reference. We take this solution as a reference, not only because it is more than converged in Fig.~\ref{fig:DOF_3D}a), but it is also visually smoother than other converged results. Accordingly, we calculate the average percent error over the superconducting domain by integrating \eqref{eq:error} for the different element orders, as summarized in Fig.~\ref{fig:DOF_3D}b).
	
	Similarly to the 2-D case, we find that linear elements are not well suited for an accurate solution in 3-D. Even for the highest number of DOF of $160,000$, the percent error is still 16~\% when using linear elements. Quadratic elements are more accurate, with the lowest error being 3~\%. Ultimately, cubic elements show the most accurate results even for numbers of DOF as low as $43,800$. Thus, cubic elements should be used in the 3-D H-formulation if the most accurate solution is desired.
	
	However, despite the accuracy of cubic elements, their computation times are significantly longer than that of lower order elements, as demonstrated in Fig.~\ref{fig:DOF_3D}c). For the highest amount of DOF simulated, cubic elements take approximately twice the amount of time required for quadratic elements and about four times longer than linear elements for the same number of DOF. Therefore, there is a compromise between accuracy and computation time for quadratic and cubic elements. Although there is not a large discrepancy between their accuracies, the computation time of quadratic elements is drastically faster than cubic elements for a given number of DOF. 
	
	\subsection{3-D H-$\phi$ formulation}
	
\begin{figure*}[tb!]
		\centering
		\includegraphics[width=0.49\linewidth]{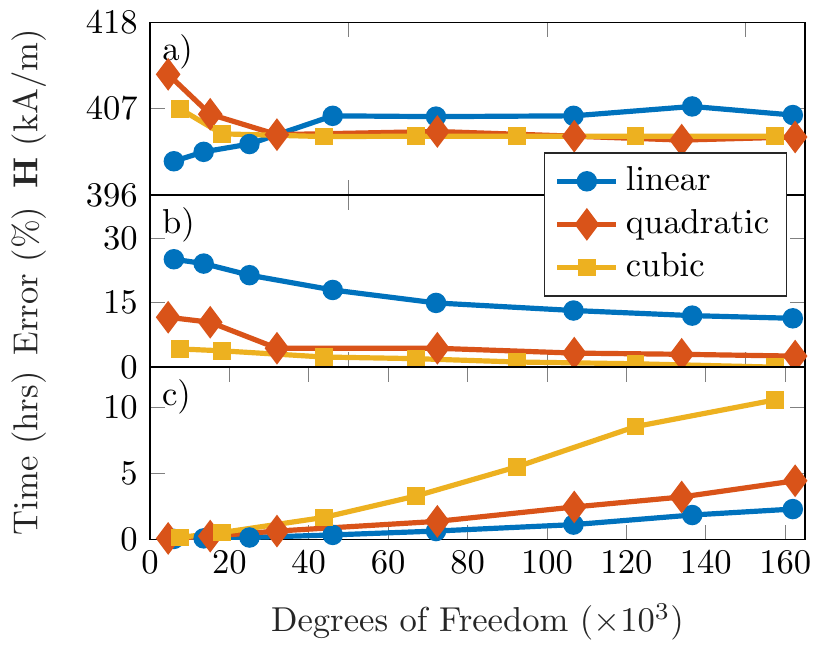}
		\includegraphics[width=0.49\linewidth]{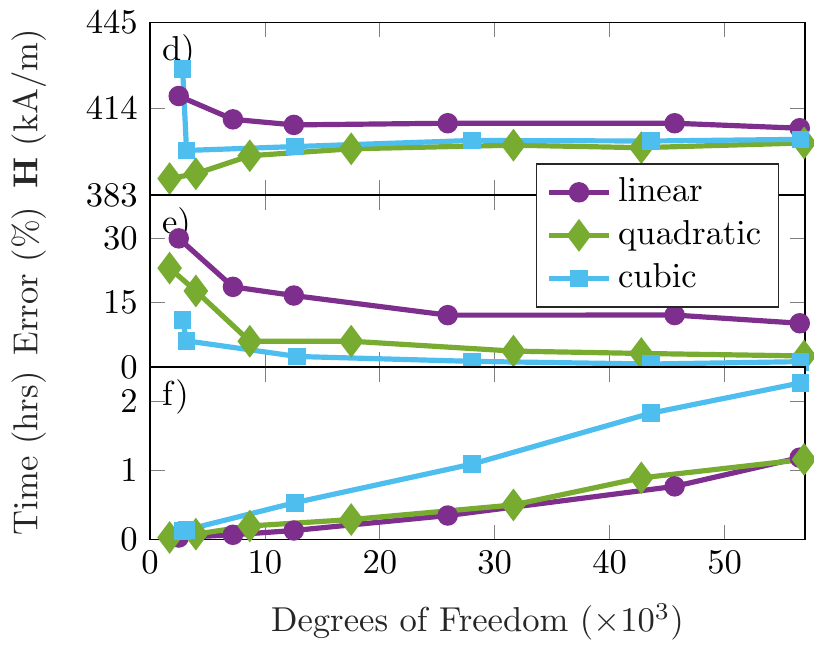}
		\caption{Comparison between 3-D simulations of the H and H-$\phi$ formulations with linear, quadratic and cubic elements. a) and d) show the convergence rates of the average field over the superconducting domain as a function of DOF in the H and H-$\phi$ formulations, respectively. b) and e) show the percent error relative to 8325 cubic curl elements simulated in the H-formulation as a function of DOF in the H and H-$\phi$ formulations, respectively. Finally, computation times of the H (c) and H-$\phi$ (f) formulations as a function of DOF are shown. A MUMPS direct solver is used for the H-formulation, while a PARDISO solver is used for the H-$\phi$ formulation.}
		\label{fig:DOF_3D}
	\end{figure*}

	We analyze the three dimensional H-$\phi$ formulation by first performing a convergence rate analysis. As shown in Fig.~\ref{fig:DOF_3D}d), cubic elements converge very rapidly to $\sim$403~kA/m, the same value as the H-formulation result. On the other hand, quadratic element results converge slower, with a convergence value slightly lower at $\sim$401~kA/m. This is surprising, considering that quadratic and quartic elements converged to approximately the same value in 2-D and the same can be said with quadratic and cubic elements in the 3-D H-formulation. Nevertheless, the difference between converged values is still less than 1~\%. Finally, linear elements converge at nearly the same rate as quadratic elements, but with a higher convergence value of $\sim$408~kA/m. Note that for a given mesh and element order, the number of DOF is reduced by a factor of over three from the H to the H-$\phi$ formulation for all simulations considered in 3-D.
	
	We proceed to calculate the percent error between the H and H-$\phi$ formulation results simulated with $157,400$~DOF and 8325~cubic elements, as illustrated in Fig.~\ref{fig:DOF_3D}e). We find that the percent error remains relatively low~($<$1.5\%) for cubic elements above 27,000~DOF. Using quadratic elements yields error values above 2.5\% for the highest number of DOF simulated, with the error remaining marginally higher than cubic elements for a given number of DOF. Linear elements are still less accurate than higher order elements, with the lowest error remaining at 10.2~\% for 56,500~DOF. Accordingly, the results are very similar to the H-formulation case: cubic elements are ideal for the most accurate solutions, with quadratic elements still providing relatively accurate results.
	
	We compare the computation times between different element orders in Fig.~\ref{fig:DOF_3D}f). Linear and quadratic elements have very similar computational efficiencies, with quadratic elements being at most 15\% slower for a given number of DOF. For the highest number of DOF simulated ($\sim$56,500), quadratic elements are $\sim$2\% faster than linear elements. On the other hand, cubic elements systematically yield much higher computation times than their lower order counterparts. Cubic elements are at least 45\% slower than linear and quadratic elements for any given number of DOF.
	
	Finally, we compare the computation times required to obtain a relative error of 3\% as a function of element order and model formulation. As shown in table~\ref{tbl:3D}, the time required to obtain a relative error of 3\% is nearly halved in the H-$\phi$ formulation when compared to the result obtained with the H-formulation using cubic elements. The computation times are nearly three times faster in the H-formulation and 0.63 times faster in the H-$\phi$ formulation for cubic elements than for quadratic elements, so we conclude that cubic elements are the best option provided by COMSOL in both formulations in 3-D. Finally, the H-$\phi$ formulation is nearly twice as fast as the H-formulation when considering cubic elements.  
	
	\begin{table}[h]
		\centering
		\caption{Computation times of the H and H-$\phi$ formulations in 3-D in order to achieve a relative error of 3\%}
		\begin{tabular}{ l  l  l }
			\toprule
			Order  & H time (hrs) & H-$\phi$ time (hrs)\\
			\midrule
			1  & $>$2.3 & $>$1.2 \\
			2  & 3.20 & 1.00 \\
			3  & 1.15 & 0.63 \\
			\bottomrule
		\end{tabular}
		\label{tbl:3D}
	\end{table}
	
We conclude the 3-D analysis with a limited comparison between mixed element orders in the H-$\phi$ formulation. Fixing the number of elements to 108 in the superconducting domain, we vary the number of elements in the air domain according to the element orders used so as to get 35,573$\pm$226 DOF. A comparison of the computation times and relative errors obtained by mixing element orders is shown in table~\ref{tbl:mix3D}. For every order of elements considered in the H physics, more accurate results are obtained if the same order is used in the $\phi$ physics. When 1:1 elements are used, the computation time is more than 40~\% faster than 1:2 and 1:3 elements and the relative error is $\sim$12\% more accurate. In addition, 2:2 elements result in slightly slower computation times, but their relative error still remains 1.44\% lower than other elements considered. Finally, 3:3 elements produce the lowest relative error out of all element orders considered, with only 1.264\% error. 3:4 elements also give relatively low error (1.273\%), but their computation times are more than 20\% slower than 3:3 elements. Using 3:2 elements yields $\sim$12\% faster simulations, but their relative error is still 0.255\% less than 3:3 elements. Additional simulations were carried out with higher amounts of DOF in order to verify if higher computation times lead to more accurate results in mixed element orders, but the equivalent order elements still result in more accurate solutions even though fewer DOF are used. Thus, we deduce that mixing element orders between the H and $\phi$ physics does not yield better results, so that 3:3 elements are still the ideal elements to use in 3-D in COMSOL in order to get the best balance between computation time and accuracy.

	\begin{table}[h]
	\centering
	\caption{Mixed element order comparison for the 3-D H-$\phi$ formulation}
	\begin{tabular}{ l  l  l  l l}
		\toprule
		H Order & $\phi$ Order & DOF & Time~(hrs) & Error~(\%) \\
		\midrule
		
		\multirow{3}{*}{1} & 1 & 35,554 & 0.442 & 12.455\\
		{}	& 2 & 35,523 & 0.723 & 22.245\\
		{}	& 3 & 35,526 & 0.990 & 24.975\\\\
		
		\multirow{3}{*}{2} & 1 & 35,525 & 0.997 & 7.136\\
		{}	& 2 & 35,533 & 1.459 & 3.132\\
		{}	& 3 & 35,595 & 1.430 & 4.564\\
		{}	& 1 & 47,129 & 1.466 & 7.053\\
		{}	& 3 & 41,353 & 1.619 & 4.555\\\\
		
		\multirow{4}{*}{3} & 2 & 35,589 & 3.104 & 1.519\\
		{}	& 3 & 35,509 & 3.533 & 1.264\\
		{}	& 4 & 35,799 & 4.295 & 1.273\\
		{}	& 2 & 48,315 & 3.780 & 1.491\\

		\bottomrule
	\end{tabular}
	\label{tbl:mix3D}
	\end{table}
	
	\section{Conclusion}
	
	In this work, we implemented the H-$\phi$ formulation in COMSOL Multiphysics in order to compare its performances with the well-documented H-formulation in the context of the magnetization of bulk superconductors. Using standard ZFC magnetization simulations in 2-D and 3-D, we studied the accuracy and computation times obtained from the different formulations with varied element orders and mesh discretizations. 
	
	By comparing simulation results of 2594 quartic elements in the H and H-$\phi$ formulations in 2-D, we found that the percent error between formulations remains below 0.3~\% at the edges of the superconducting domain, where the field varies more drastically. The percent error remained below 0.01~\% for fields far from the bulk, showing that the formulations give nearly equivalent results even though different element types were used.  
	
	We identified the ideal element order to be the highest order implementable in COMSOL Multiphysics, regardless of formulation or dimension. Accordingly, the ideal element order in COMSOL is quartic in 2-D and cubic in 3-D for both formulations. The choice of element orders is clear in 2-D: both the accuracies and computation times are improved for quartic elements. In order to obtain a relative error of 0.5\%, the computation times were found to be more than 1.5 times quicker using quartic elements than quadratic elements. In 3-D, cubic elements provide greater accuracy than quadratic elements for a given number of DOF, but their computation times are much higher with their accuracy being merely slightly better. Nevertheless, we find that cubic elements used in the H-$\phi$ formulation still solve virtually twice as fast as cubic elements in the H-formulation when comparing the computation times required for an error of 3\%. 
	
	By conducting a limited analysis of mixing Lagrange and curl element orders in the H-$\phi$ formulation, we found that using the same orders for both elements offers a better balance between computation time and accuracy. 
		
	Further work on more sophisticated simulations representing, for instance, electrical machines or other applications could be realized in order to generalize the observations presented in this work.
	
	\section{Acknowledgements}
	
	The authors would like to acknowledge Bruno Alves for interesting discussions concerning the finite element method and Can Superconductors for providing the field dependent critical current density data. 
	
	This work was supported by the Fonds de recherche du Qu\'ebec --- Nature et Technologies (FRQNT) and TransMedTech Institute and its main funding partner, the Canada First Research Excellence Fund.

	
	%

	\appendix[COMSOL implementation of the H-$\phi$ formulation]
	
In this Appendix, we describe the implementation of the H-$\phi$ formulation in COMSOL. The model used in this work will be available on the HTS modelling website \cite{htsmodeling}. 

Similarly to the H-formulation, there are two equivalent ways of implementing the H-$\phi$ formulation: by defining our own PDEs or using the predefined COMSOL \textit{Magnetic Field Formulation} (MFH) and \textit{Magnetic Field No Currents} (MFNC) modules. We describe the more general 3-D implementation, the 2-D case easily follows.

The PDE implementation follows from the equations laid out in Sec.~\ref{sec:H-phi}. The \textbf{H} physics is implemented using the \textit{General Form PDE} physics to introduce \eqref{eq:H} in the superconducting domain, as done in the regular H-formulation. We introduce the $\phi$ physics in the non-conducting domain with a \textit{Weak Form PDE} node and implement the first term of \eqref{eq:weakphi}, corresponding to the Lagrange equation. In COMSOL notation, this is given in 3-D by:
\begin{equation}
\textnormal{-ux*test(ux)-uy*test(uy)-uz*test(uz)},
\end{equation}
where u is the dependent variable of the $\phi$ physics, ui is the derivative of u with respect to i, and test(u) is the test function defined by COMSOL. By defining this expression in the \textit{Weak Form PDE} node, COMSOL automatically takes the integral of the expression over the selected domain and sets it equal to zero. The Laplace equation can also be implemented using the \textit{Coefficient Form PDE} module.

Finally, we apply the background field by using a regular \textit{Dirichlet Boundary Condition} node, keeping in mind that the negative of the gradient of the inserted expression generates the applied background magnetic field.

In order to couple the physics together, we use the procedure outlined in Sec.~\ref{sec:couple}. First, the tangential components of the \textbf{H}-field are constrained to the tangential components of the \textbf{h}-field by using a \textit{Constraint} node in the \textbf{H} physics with the expressions
\begin{align}
\textnormal{tHx+uTx=0}\\
\textnormal{tHy+uTy=0}\\
\textnormal{tHz+uTz=0}
\end{align}
introduced in the three boxes supplied. Here, tHi represents the tangential component of the \textbf{H}-field in the i-direction and uTi represents the tangential derivative of u in the i direction. The constraint settings must be set to \textit{Current physics (internally symmetric)} in order to get a unidirectional constraint and not overconstrain the \textbf{h}-field. 

Equating the normal components of the magnetic fields can easily be done with the \textit{Flux/Source} node in the $\phi$ physics. The equation defined under this node is given by $-\mathbf{n}\cdot \nabla u=g$, where g is the boundary source term. We therefore set $\textnormal{g=nx*Hx+ny*Hy+nz*Hz}$, keeping in mind that the \textbf{h}-field is given by $-\nabla$u, so that g should be positive. Here, nx, ny, and nz are the components of the vector normal to the surface. We should also pay careful attention to the normal vectors used, since the normal of one domain is equal to the negative of the normal of the other. In this case, simply typing nx, ny and nz supplies the normal vector on $\Gamma_{SC}$ of the air domain, since we applied the \textit{Flux/Source} node in the $\phi$ physics.

The H-$\phi$ formulation can effortlessly be implemented with built-in MFH and MFNC modules. Although less control is given on defined variables in this case, the predefined physics modules make it very easy to carry out simulations without the trouble of defining all necessary variables. However, for reasons unknown to the authors, quartic elements are unavailable in the 2-D MFH module. Also, the reference frame cannot be changed in predefined modules, which is important for simulations with a moving mesh.

Let's start with the \textbf{H} physics, imposed using the MFH module. The superconducting physics is simply imposed by using a nonlinear resistivity in the \textit{Faraday's Law} node. We then couple to the MFNC physics by using a \textit{Magnetic Field} node, which generates a magnetic field at the boundary of the superconductor. The input is simply (mfnc.Hx,mfnc.Hy,mfnc.Hz), where mfnc.H is the magnetic field calculated in the MFNC module. COMSOL automatically equates the tangential components in this formulation (as can be seen in the \textit{equation view}), since we are using edge elements. Again, we set the constraint settings to \textit{Current physics (internally symmetric)} in order to get a unidirectional constraint.

Finally, the $\phi$ physics is implemented using the MFNC module. There are two ways of applying the background field, either by applying it at the boundary using a \textit{Magnetic Flux Density} node, or by solving for the reduced field in the MFNC node and imposing a background field. In the latter case, an \textit{External Magnetic Flux Density} node needs to be applied at the domain boundary in order to generate the field. However, no significant difference has been observed between the two methods of imposing the background field. The coupling between the MFNC and MFH physics is done with a \textit{Magnetic Flux Density} node with (mfh.Bx, mfh.By, mfh.Bz) as input on $\Gamma_{SC}$, where mfh.B is the magnetic flux density calculated using the MFH physics. 	

The magnetic scalar potential must be gauged in order to obtain a unique value for $\phi$ when applying the magnetic field using a \textit{Magnetic Flux Density} node, since this node only specifies the flux of the magnetic flux density. In the simulations considered in this work, $\phi$ is gauged by the application of the field with the Dirichlet boundary conditions. In addition, Gauss' law is automatically imposed in COMSOL's MFH module. We found that this is not necessary for time-dependent simulations and leads to slower computation times.
	
	\ifCLASSOPTIONcaptionsoff
	\newpage
	\fi

	
	
	%
\begingroup
\raggedright
\bibliography{H_phi_IEEE_arxiv}
\bibliographystyle{IEEEtran}
\endgroup
	
	%
	
	\begin{IEEEbiographynophoto}{Alexandre Arsenault}
		received a B.Sc. in physics from McGill University, Montr\'eal, QC, Canada, in 2016. He also received a M.Sc. in physics from McMaster University, Hamilton, ON, Canada, in 2018. He is currently pursuing a Ph.D. degree in biomedical engineering at Polytechnique Montr\'eal under the supervision of Dr. Fr\'ed\'eric Sirois. His research interests include the characterization and simulation of bulk high-temperature superconductors.
	\end{IEEEbiographynophoto}

\begin{IEEEbiographynophoto}{Fr\'ed\'eric Sirois}(S'96--M'05--SM'07)
	received the B.Eng. degree in electrical engineering from Universit\'e de Sherbrooke, Sherbrooke, QC, Canada, in 1997, and the Ph.D. degree in electrical engineering from Polytechnique Montr\'eal, Montr\'eal, QC, Canada, in 2003.
	From 1998 to 2002, he was affiliated as a Ph.D. scholar with the Hydro-Qu\'ebec's Research Institute (IREQ), where he was a Research Engineer from 2003 to 2005. In 2005, he joined Polytechnique Montr\'eal, where he is currently Full Professor. His main research interests are i) the characterization and modeling of electric and magnetic properties of materials, ii) modeling and design of electromagnetic and superconducting devices, and iii) integration studies of superconducting equipment in power systems. He is a regular reviewer for several international journals and conferences.
\end{IEEEbiographynophoto}
	
	\begin{IEEEbiographynophoto}{Francesco Grilli}
		
		received the M.S. degree in Physics from the University of Genoa, Italy, in 1998, the Ph.D. degree in Technical Sciences from the \'Ecole Polytechnique F\'ed\'erale de Lausanne, Switzerland, in 2004, and the Habilitation in superconductivity for energy applications from the Karlsruhe Institute of Technology, Germany, in 2017.
		
		From 2004 to 2007, he was a Postdoctoral Researcher with the Los Alamos National Laboratory, NM, USA, and from 2007 to 2009, with Polytechnique Montr\'eal, QC, Canada. Since 2009, he has been with the Karlsruhe Institute of Technology, Germany, where he is currently the leader of the group ``AC Losses in High-Temperature Superconductors.'' His main research interests include the 2D and 3D modeling of high-temperature superconductors and the characterization of their properties.
		
		Dr. Grilli was the recipient of the 2008 and 2014 Van Duzer Prize for best contributed non-conference paper published in the IEEE TRANSACTIONS ON APPLIED SUPERCONDUCTIVITY and of the 2011 Dr. Meyer-Struckmann Science Prize for his work on numerical modeling of superconductors.
		
	\end{IEEEbiographynophoto}

	
	

\end{document}